\documentclass[reprint,nofootinbib,superscriptaddress,amsmath,amssymb,aps]{revtex4-1}
\usepackage{graphicx}
\usepackage{rotating}
\usepackage{dcolumn}
\usepackage{bm}
\usepackage{color}
\usepackage{mathptmx, textcomp}
\usepackage[latin1]{inputenc}
\usepackage{braket}
\usepackage[colorlinks,linkcolor=magenta,anchorcolor=cyan,citecolor=blue]{hyperref}
\usepackage[multiple]{footmisc}

\def\be{\begin{equation}}
\def\ee{\end{equation}}
\def\ba{\begin{eqnarray}}
\def\ea{\end{eqnarray}}

\newcommand{\eq}[1]{(\ref{#1})}

       \def\a {\alpha} \def\s {\sigma} \def\d {\delta}  \def\g {\gamma}   \def\k {\kappa}   \def\x {\xi} \def\c {\chi} \def\b {\beta}      
              \def\grad{\nabla}\def\.{\cdot}
\def\math {\mathcal}
\begin{document}

\title{Black Hole Zeroth Law in Horndeski Gravity}
\author{Aofei Sang}
\email{202021140021@mail.bnu.edu.cn}
\affiliation{College of Education for the Future, Beijing Normal University,
Zhuhai, 519087, China}
\affiliation{Department of Physics, Beijing Normal University, Beijing 100875, China\label{addr2}}
\author{Jie Jiang}
\email{Corresponding author. jiejiang@mail.bnu.edu.cn}
\affiliation{College of Education for the Future, Beijing Normal University,
Zhuhai, 519087, China}
\affiliation{Department of Physics, Beijing Normal University, Beijing 100875, China\label{addr2}}
\date{\today}
\begin{abstract}
  The four laws of black hole mechanics have been put forward for a long time. However, the zeroth law, which states that the surface gravity of a stationary black hole is a constant on the event horizon, still lacks universal proof in various modified gravitational theories. In this paper, we study the zeroth law {in a special Horndeski gravity}, which is an interesting gravitational theory with a nonminimally coupled scalar field. After assuming that {the nonminimally coupled scalar field has the same symmetries with the spacetime,} the minimally coupled matter fields satisfy the dominant energy condition and the Horndeski gravity has a smooth limit to Einstein gravity when the coupling constant approaches zero, we prove the zeroth law based on the gravitational equation in Horndeski gravity without any assumption to the spacetime symmetries.
\end{abstract}
\maketitle
\section{Introduction}

Black hole is an object predicted by general relativity and its existence is ensured by the singularity theorem\cite{Penrose:1969pc}. In the 1970s, the four laws of black hole mechanics are proposed\cite{Bardeen:1973gs}. Later, Hawking radiation, using the method of the quantum field theory in curved spacetime, gives a relationship between the surface gravity $\kappa$ and the temperature of a stationary black horizon and therefore indicates that black hole can be regarded as a thermodynamic system\cite{Hawking:1975vcx}. Since then, the four laws of black hole mechanics and other consequent thermodynamic properties of the black hole arose wide attention.

The zeroth law of black hole mechanics, which states that the surface gravity $\kappa$ (temperature) of a stationary black hole is a constant on the event horizon, is proved in Einstein gravity in 1970s\cite{Bardeen:1973gs}. In Einstein gravity, guaranteed by the rigidity theorem\cite{Hawking:1971vc}, the event horizon must be a Killing horizon of a black hole in asymptotic flat stationary spacetime. Then, using the Einstein equation and the dominant energy condition, one can easily find that the derivative of $\kappa$ in all directions along the horizon is zero\cite{Bardeen:1973gs}. However, this proof failed in other gravitational theories since the Einstein equation is used in the process of proof. Later, a proof without the equation of motion for the gravitational theory is given by Racz and Wald\cite{Racz:1995nh}. They assume that the black hole is static or stationary-axisymmetric with `$t-\phi$' reflection isometry and find that in this case the zeroth law always holds. This proof is independent of the concrete gravitational theories, however, in contrast to the argument in the previous proof, this argument makes quite strong assumptions about the symmetry of black hole spacetime. In short, there is still a lacks general proof of the black hole zeroth law\cite{Sarkar:2012wy,Ghosh:2020dkk}.

General relativity has been a great success in describing gravity, especially in explaining phenomena in the solar system, but many observations of the universe indicate that there is still a lot to be discussed for this theory. For example, general relativity tells us that the universe is slowing down, which contradicts the observational result that the universe is accelerating. Many other cosmological problems, such as dark matter, dark energy, and the early origins and the evolution of the universe, also remain to be solved\cite{dePireySaintAlby:2017lwc,Planck:2015fie,Bertone:2004pz,Peebles:2002gy,Alpher:1949sef}. One way to solve these problems is to introduce the scalar field. The Horndeski gravity considered in this paper, first proposed in the 1970s, is a general second-order scalar-tensor theory\cite{Horndeski:1974wa}. {The full action of this kind of theory is given in \cite{Deffayet:2011gz, Kobayashi:2011nu}.} Recently, this gravitational theory received renewed attention for the possibility to solve the cosmological constant problem\cite{Charmousis:2011bf} and other interesting properties\cite{Nicolis:2008in,Saridakis:2010mf,Maselli:2016gxk}. Many thermodynamic properties of stationary black holes in Horndeski gravity are also discussed recently such as the black hole first law, second law and etc.\cite{Feng:2015oea,Miao:2016aol,Feng:2015wvb,Hu:2018qsy,Hajian:2020dcq,Wang:2020svl}. However, there is no general proof for the black hole zeroth law in Horndeski gravity and a natural question is whether the zeroth law holds generally in Horndeski gravity. In this paper, we explored this question in the stationary black holes of {a special Horndeski gravity}. Then, with natural assumptions that {the nonminimally coupled scalar field shares the same symmetries with the spacetime,} the Horndeski gravity approach to general relativity smoothly when the coupling constant approach to zero and the matter field satisfies the dominant energy condition, we would like to prove that the surface gravity is constant in the whole event horizon, i.e. the zeroth law of black hole mechanics established in Horndeski gravity. This result is general in Horndeski gravity since there is no extra symmetry is assumed in the process of proof.

This paper is organized as follows. In sec. \ref{sec2}, we introduce some properties of the Killing horizon, which will be used for further proof and review the proof of the zeroth law in Einstein gravity. Then, we introduce the Horndeski gravity and gives the equation of motion of this theory in sec. \ref{sec3}. The main calculation of the proof is shown in sec. \ref{sec4}. {Finally, in sec. \ref{sec5}, we give a conclusion and prospects the later work.}

\section{Geometry of stationary Killing horizon}\label{sec2}
In this section, we would like to introduce some related properties of the stationary event horizon which is also the Killing horizon in the $n$-dimensional stationary black holes.

A null hypersurface $\math{H}$ is called a Killing horizon if there exists a Killing vector field $\xi^a$ which is orthogonal to the null hypersurface $\math{H}$. The surface gravity $\kappa$ of the Killing horizon is defined by
\be\begin{aligned}\label{kappa}
\xi^a\nabla_a\xi^b=\kappa \xi^b\,.
\end{aligned}\ee
To prove the zeroth law, we need to check that whether the surface gravity $\kappa$ is a constant on the hypersurface $\math{H}$. Choose $C$ is a cross-section of the event horizon and another cross-section in $\math{H}$ is generated by the Killing vector field $\x^a$. For calculation convenient, we can choose a basis $\{\xi^a,l^a,(e_i)^a\}$ satisfying
\be\begin{aligned}
&\xi^a l_a=-1\,,\quad l^a l_a=0\,,\\
\xi_a (e_i)^a=&0\,,\quad l_a (e_i)^a=0\,,\quad (e_i)^a (e_j)_a=\d_{ij}\,.
\end{aligned}\ee
on $\math{H}$ and $(e_i)^a$ is tangent vectors of the cross-section $C$. Then, we can express the metric on the horizon as
\be\begin{aligned}\label{metric}
g_{ab}=-\xi_a l_b-\xi_b l_a+\gamma_{ab}\,,
\end{aligned}\ee
where $\gamma_{ab}=\d_{ij}(e^i)_a(e^j)_b$ is the induced metric of the cross-section $C$. Using this basis and the definition of $\kappa$, we can get
\be\begin{aligned}
\kappa=l_b\xi^a \nabla_a \xi^b\,.
\end{aligned}\ee
The Killing horizon of a stationary black hole can be characterized by the vanishing shear $\s_{ab}$ and expansion $\vartheta$ of the horizon, which are defined by
\ba\begin{aligned}
\vartheta=\frac{1}{2}\g^{ab}\math{L}_\x \g_{ab}\,,\quad\quad \s_{ab}=\frac{1}{2}\math{L}_\x \g_{ab}-\frac{\vartheta}{n-2}\g_{ab}\,.
\end{aligned}\ea
Using the fact that $\x^a$ is a Killing vector field, it is easy to see
\be\begin{aligned}\label{alongxi}
\xi^a\nabla_a\kappa=0\,.
\end{aligned}\ee
We can note that the derivation of Eq. \eqref{alongxi} did not use the equation of motion and therefore it is true for any gravitational theories. To prove the constancy of surface gravity, we only need to show $\k$ vanishes along the transverse directions, i.e., $D_a \k=\g_{a}{}^b\grad_b \k=0$. Here $D_a$ is the induced covariant derivative operator on the cross-section $C$ which is defined by
\ba\begin{aligned}
D_a T_{bcd\cdots} := \g_a{}^{a_1}\g_b{}^{b_1}\g_c{}^{c_1}\cdots \grad_{a_1} T_{b_1c_1d_1\cdots}
\end{aligned}\ea
for any tensor $T_{bcd\cdots}$ on $C$.

Using the Gauss-Codazzi equation \cite{Poisson2004,Vega:2011ue} and considering the fact that expansion and shear vanish on the entire horizon, we can get
\be\begin{aligned}\label{Rabcd}
&R_{abcd}\xi^a (e_i)^b \xi^c (e_j)^d=0\,,\\
&R_{abcd}\xi^a (e_i)^b (e_j)^c (e_k)^d=0\,,\\
&R_{ab}\xi^a\xi^b=0
\end{aligned}\ee
on the horizon $\math{H}$. Using these conditions, it is straightforward to get
\be\begin{aligned}\label{alongei}
D_b\kappa=R_{aecd}\xi^a l^e \gamma_b{}^c \xi^d=-R_{ac}\xi^a\gamma_b{}^c\,.
\end{aligned}\ee
For the Einstein gravity, the equation of motion gives
\be\begin{aligned}\label{Einsteineom}
G_{ab}=R_{ab}-\frac{1}{2}g_{ab}R=8\pi T_{ab}\,.
\end{aligned}\ee
Thus, the derivative of $\kappa$ in the spatial direction becomes
\be\begin{aligned}
D_b\kappa=-8\pi T_{ac}\xi^a\gamma_b{}^c\,.
\end{aligned}\ee
Moreover, combining \eqref{Rabcd} and \eqref{Einsteineom} also gives $T_{ab}\xi^a\xi^b|_{\math{H}}=0$, which means $T_{ab}\xi^b$ can be written as linear combination of $\xi_a$ and $(e_i)_a$ on $\math{H}$. If the matter field satisfies the dominant energy condition, $T_{ab}\xi^b$ must be timelike or null. Therefore, $T_{ab}\xi^b\propto \xi_a$ on $\math{H}$, which implies $T_{ac}\x^a\g_b{}^c=0$. These imply that $D_a\kappa=0$ in Einstein gravity, and therefore $\kappa$ is a constant on the horizon.

\section{The Horndeski Gravitational theory}\label{sec3}
In this paper, {we would like to consider the zeroth law in the Horndeski gravity with non-minimally coupled scalar field $\chi$ and linear curvature term. A full Horndeski action can be found in Refs. \cite{Deffayet:2011gz,Kobayashi:2011nu}. In this paper, we would like to consider a special case of the full theory for simplification. We take $K=\beta X$, $G_3=0$, $G_4=G_5=1$ in Ref. \cite{Kobayashi:2011nu}. Then, the action of a simplified $n$-dimension Horndeski gravitational theory can be written as\cite{Wang:2020svl}}
\be\begin{aligned}
I=I_{\text{Horn}}+I_{\text{matter}}\,,
\end{aligned}\ee
where $I_{\text{matter}}$ is the action of the minimally coupled matter fields, and $I_{\text{Horn}}$ is given by
\be\begin{aligned}
I_\text{Horn}=\frac{1}{16\pi}\int d^{n}x \sqrt{-g}\left[R-\frac{1}{2}(\beta g^{ab}-\alpha G^{ab})\nabla_a\chi\nabla_b\chi\right]\,,
\end{aligned}\ee
where $G_{ab}$ is the Einstein tensor and $\alpha$, $\beta$ are some coupling constants. The equation of motion is given by\cite{Wang:2020svl}
\be\begin{aligned}\label{horneom}
H_{ab}=8\pi T_{ab}\,,
\end{aligned}\ee
in which
\ba\begin{aligned}
H_{ab}=&G_{ab}-\frac{\alpha}{4}G_{ab}(\nabla\chi)^2-\frac{\alpha}{4}g_{ab}A-\frac{\alpha}{2}K_{ab}-\b H^{(\b)}_{ab}
\end{aligned}\ea
with
\be\begin{aligned}
&A=(\nabla^c\chi\nabla^d\chi)(\nabla_c\chi\nabla_d\chi)-(\nabla^2\chi)^2+2R^{cd}\nabla_c\chi\nabla_d\chi\,,\\
&K_{ab}=\frac{R}{2}\nabla_a\chi\nabla_b\chi-2\nabla_c\chi\nabla_{(a}\chi R^c_{b)}+\nabla_a\nabla_b\chi\nabla^2\chi\\
&\quad\quad-R_{acbd}\nabla^c\chi\nabla^d\chi-\nabla_a\nabla^c\chi\nabla_b\nabla_c\chi\,,\\
&H^{(\b)}_{ab}=\frac{1}{2}\nabla_a\chi\nabla_b\chi-\frac{1}{4}g_{ab}\nabla^c\chi\nabla_c\chi\,.
\end{aligned}\ee
Here $T_{ab}$ represents the stress-energy tensor of the minimally coupled matter fields. In this paper, we only assume these minimally coupled matter fields satisfy the dominant energy condition, i.e., $\xi^a T_{ab}$ must be non-spatial.

\section{the black hole zeroth law in horndeski gravity}\label{sec4}
In this section, we would like to investigate the zeroth law in Horndeski gravity. For this purpose, we divide the main calculation into three parts. In the first part, we calculate the contraction of the equation of motion and $\xi^a\xi^b$. In the second part, we calculate the contraction of the equation of motion and $\xi^a(e_i)^b$. In the third part, we analyze the identity we have found and gives a discussion.
\subsection{Contract with $\xi^a\xi^b$}
We would like to contract $\xi^a\xi^b$ both side of Eq. \eqref{horneom} in this subsection. Considering that $\xi^a$ is a null vector on the horizon, the contraction of $H_{ab}$ and $\x^a\x^b$ gives
\be\begin{aligned}\label{contractxi}
H_{ab}\xi^a\xi^b&=\left[1-\frac{\alpha}{4}(\nabla\chi)^2\right]R_{ab}\xi^a\xi^b\\
&-\frac{\alpha}{2}K_{ab}\xi^a\xi^b-\frac{\b}{2}\x^a\grad_a\c\x^b\grad_b\c\,.
\end{aligned}\ee
{Note that $\xi^a$ is a Killing vector field. In this paper, we would like to consider a special solution of the Horndeski gravity where the non-minimally coupled scalar field $\chi$ shares the same symmetries with the spacetime, i.e., we have $\math{L}_{\xi}\chi=\xi^a\nabla_a\chi=0$} and therefore the last term of the above expression vanishes. Together with the third equation in Eq. \eqref{Rabcd}, the first term of Eq. \eq{contractxi} also vanishes. Then, Eq. \eq{contractxi} further simplifies to
\be\begin{aligned}\label{Kxixi}
H_{ab}\x^a\x^b=-\frac{\alpha}{2}K_{ab}\xi^a\xi^b\,.
\end{aligned}\ee
Again using the Killing condition $\x^a\grad_a\c=0$, the lhs of
Eq. \eqref{Kxixi} becomes
\be\begin{aligned}\label{Kxixi1}
K_{ab}\xi^a\xi^b=&\xi^a\xi^b\nabla_a\nabla_b\chi\nabla^2\chi-\x^a\x^bR_{acbd}\nabla^c\chi\nabla^d\chi\\
&-\x^a\x^b\nabla_a\nabla^c\chi\nabla_b\nabla_c\chi\,.
\end{aligned}\ee
The first term vanishes on the horizon since
\be\begin{aligned}\label{first}
\xi^a\xi^b\nabla_a\nabla_b\chi&=\xi^a\nabla_a(\xi^b\nabla_b\chi)-(\xi^a\nabla_a\xi^b)\nabla_b\chi\\
&=\math{L}_\x^2\c-\kappa\xi^b\nabla_b\chi=0\,,
\end{aligned}\ee
where we used the definition of the surface gravity and the fact that $\xi^a$ is the Killing vector field. Then, using the decomposition of the metric \eqref{metric} and considering the symmetries of the Riemann tensor, the second term induces to
\be\begin{aligned}
&\xi^a\xi^b R_{acbd}\nabla^c\chi\nabla^d\chi=\xi^a\xi^b R_{acbd}g^{ce}g^{df}\nabla_e\chi\nabla_f\chi\\
&=R_{acbd}\xi^a\xi^bD^c\chi D^d\chi-2\x^a\x^bR_{acbd}l^c \x^e\grad_e\c D^d\c\\
&\quad +\x^a\x^b l^c l^d R_{acbd} \x^e\grad_e \c\x^f\grad_f\c  \,.
\end{aligned}\ee
According to \eqref{Rabcd}, $R_{aebf}\gamma_c{}^e\gamma_d{}^f\xi^a\xi^b$ vanishes on the horizon. Together with the Killing condition $\x^a\grad_a\c=0$, it is easy to get
\ba
\xi^a\xi^b R_{acbd}\nabla^c\chi\nabla^d\chi=0
\ea
on the horizon.

Similarly, using Eq. \eqref{metric}, we can expand the third term of Eq. \eq{Kxixi1} as
\be\begin{aligned}
&\xi^a\xi^b \nabla_a \nabla^c \chi \nabla_b \nabla_c \chi=\xi^a\xi^b g^{cd} \nabla_a \nabla_d \chi \nabla_b \nabla_c \chi\\
&=\xi^a\xi^b\gamma^{cd} \nabla_a \nabla_d \chi \nabla_b \nabla_c \chi-2\xi^a\xi^b l^{c}\xi^{d} \nabla_a \nabla_d \chi \nabla_b \nabla_c \chi\\
&=\xi^a\xi^b\gamma^{cd} \nabla_a \nabla_d \chi \nabla_b \nabla_c \chi\,,
\end{aligned}\ee
where we have used Eq. \eqref{first} in the last step. As for the remaining part, we consider
\ba\begin{aligned}\label{contractexi}
  \xi^a\gamma_c{}^d\nabla_d\nabla_a\chi&=\gamma_c{}^d\nabla_d(\x^a\nabla_a\chi)-(\gamma_c{}^d\nabla_d\xi^a)\nabla_a\chi\\
  &=-(\gamma_c{}^d\nabla_d\xi^a)\nabla_a\chi\,.
  \end{aligned}\ea
Then, it can be expanded as
\be\begin{aligned}
\xi^a\gamma_c{}^d\nabla_d\nabla_a\chi&=-(\delta_c{}^d+\xi_c l^d+\xi^d l_c)\nabla_d\xi^a\nabla_a\chi\\
%&=-\nabla^a\chi(\nabla_c\xi_a+l^d\xi_c\nabla_d\xi_a+l_c\xi^d\nabla_d\xi_a)
&=-\nabla^a\chi(\nabla_c\xi_a+\kappa l_c\xi_a+l^b\xi_c\nabla_b\xi_a)\,.
\end{aligned}\ee
For the Killing horizon generated by $\xi^a$, we have $\xi^a$ is hypersurface orthogonal, i.e., $\xi^a$ satisfies
\be\begin{aligned}
\xi_c\nabla_b\xi_a=\xi_b\nabla_c\xi_a+\xi_a\nabla_b\xi_c\,
\end{aligned}\ee
on the horizon. After taking this condition into account, we can find
\be\begin{aligned}\label{exigrad}
  \xi^a\gamma_c{}^d\nabla_d\nabla_a\chi&=-\nabla^a\chi(\nabla_c\xi_a+\kappa l_c\xi_a+l^b\xi_c\nabla_b\xi_a)\\
  &=-\nabla^a\chi(\nabla_c\xi_a+\kappa l_c\xi_a+l^b\xi_b\nabla_c\xi_a+l^b\xi_a\nabla_b\xi_c)\\
  &=-\nabla^a\chi[\nabla_c\xi_a-\nabla_c\xi_a+\xi_a(\kappa l_c+l^b\nabla_b\xi_c)]\\
  &=0\,.
\end{aligned}\ee
Finally, we find the three terms in Eq. \eqref{Kxixi1} all vanish on the horizon and therefore we have{
\be\begin{aligned}\label{horizonT}
8\pi T_{ab}\xi^a\xi^b=H_{ab}\xi^a\xi^b=0
\end{aligned}\ee
on the horizon $\math{H}$.} {With a similar discussion in sec. \ref{sec2}, we assume the minimally coupled matter field satisfies the dominant energy condition. The dominant energy condition gives that for any future-directed, timelike vector $Z^a$, $-T^a_b Z^b$ is further directed timelike or null vector. Therefore, for the future-directed null vector $\xi^a$, there must be $-T_{ab}\xi^a Z^b \leq 0$ for any future-directed timelike vector $Z^a$, which means $T_{ab}\xi^a$ cannot be spacelike. In addition, considering the result \eqref{horizonT}, we can written $T_{ab}\xi^b$ as linear combination of $\xi_a$ and $(e_i)_a$ on $\math{H}$. Combining with $T_{ab}\xi^b$ must be timelike or null, it is easy to find $T_{ab}\xi^b\propto \xi_a$ on $\math{H}$, which implies}
\ba\label{domT}
T_{ab}\xi^a\gamma_c{}^b=0
\ea
on the horizon $\math{H}$ in the Horndeski gravity.

\subsection{Contract with $\xi^a\gamma_c{}^b$}

In this subsection, we would like to find the corresponding relation of $D_b\k$ using the equation of motion in Horndeski gravity to prove the zeroth law. To do this, we contract $\xi^a\gamma_c{}^b$ both side of \eqref{horneom}. Since $\xi^a$ is orthogonal to the cross-section and the Killing condition $\xi^a\nabla_a\chi=0$, the contraction $H_{ac}\x^a\g_b{}^c$ gives
\be\begin{aligned}\label{contractei}
   H_{ab}\xi^a\gamma_c{}^b&=\left[1-\frac{\alpha}{4}(\nabla\chi)^2\right]R_{ab}\xi^a\gamma_c{}^b-\frac{\alpha}{2}K_{ab}\xi^a\gamma_c{}^b\\
  &=-\left[1-\frac{\alpha}{4}(D\chi)^2\right]D_c\k-\frac{\alpha}{2}K_{ab}\xi^a\gamma_c{}^b\,.
\end{aligned}\ee

Then, we would like to calculate the remaining five terms in $K_{ab}\xi^a\gamma_c{}^b$. The first term is vanishing since $\xi^a\nabla_a\chi=0$. For the second term, it can be written as
\be\begin{aligned}\label{second2}
&-2\xi^a\gamma_c{}^b\nabla_d\chi\nabla_{(a}\chi R^d_{b)}=-\xi^a\gamma_c{}^b R_{ad}\nabla^d\chi\nabla_{b}\chi\\
&=-(-l^{(a}\xi^{e)}+\gamma^{ae})(-l^{(d}\xi^{f)}+\gamma^{df})\xi_e\gamma_c{}^b R_{ad}\nabla_f\chi\nabla_{b}\chi\,.
\end{aligned}\ee
After expanding Eq. \eqref{second2} and considering the Killing condition, the orthogonality as well as $R_{ab}\xi^a\xi^b=0$, we can find the only term left is
\be\begin{aligned}
-2\xi^a\gamma_c{}^b\nabla_d\chi\nabla_{(a}\chi R^d_{b)}&=-R_{ad}\gamma_c{}^b\gamma^{df}\xi^a\nabla_b\chi\nabla_f\chi\\
&=D^a\c D_c\c D_a\k\,.
\end{aligned}\ee
For the third term, using what we have found in Eq. \eqref{exigrad}, it is easy to find that this term vanishes. The fourth term can be calculated using the same method. It is not hard to find
\be\begin{aligned}
  &-\xi^a\gamma_c{}^b R_{aebd}\nabla^e\chi\nabla^d\chi\\
  %=-g^{ef} g^{dg}\xi^a\gamma_c{}^b R_{aebd} \nabla_f\chi\nabla_g\chi\\
  &=-(-l^{(e}\xi^{f)}+\gamma^{ef}) (-l^{(d}\xi^{g)}+\gamma^{dg})\xi^a\gamma_c{}^b R_{aebd} \nabla_f\chi\nabla_g\chi\\
  &=R_{aedf}\gamma_g{}^e\gamma_b{}^f\gamma_c{}^d\xi^a\nabla^g\nabla^b\chi+R_{gebf}l^a\gamma_c{}^e\gamma_d{}^f\xi^g\xi^b\nabla_a\chi\nabla^d\chi
\end{aligned}\ee
after using that $\xi^a$ is the Killing vector field. Then, considering Eq. \eqref{Rabcd}, we can find the above two terms all vanish on the horizon, i.e., we have
\be\begin{aligned}
  -\xi^a\gamma_c{}^b R_{aebd}\nabla^e\chi\nabla^d\chi=0\,.
\end{aligned}\ee
For the fifth term,
%\be\begin{aligned}
%  -\xi^a\gamma_c{}^b\nabla_a\nabla^d\chi\nabla_b\nabla_d\chi=-g_{aa1}g_{bb1}g_{dd1}\xi^a\gamma_c{}^b\nabla^{a1}\nabla^d\chi\nabla^{b1}\nabla^{d1}\chi\,.
%\end{aligned}\ee
we do the same as before. Using the expression of the metric and considering the orthogonal relation between the basis, we can get
\be\begin{aligned}
  &-\xi^a\gamma_c{}^b\nabla_a\nabla^d\chi\nabla_b\nabla_d\chi=l_d\gamma_{bf}\gamma_c{}^b\xi_a\xi_e\nabla^e\nabla^d\chi\nabla^f\nabla^a\chi\\
  &+l_a\gamma_{bf}\gamma_c{}^b\xi_d\xi_e\nabla^e\nabla^d\chi\nabla^f\nabla^a\chi-\gamma_{bf}\gamma_c{}^b\gamma_{da}\xi_e\nabla^e\nabla^d\chi\nabla^f\nabla^a\chi\,.
\end{aligned}\ee
Then, using Eqs. \eqref{first} and \eqref{exigrad}, it is obvious that
\be\begin{aligned}
  -\xi^a\gamma_c{}^b\nabla_a\nabla^d\chi\nabla_b\nabla_d\chi=0\,.
\end{aligned}\ee

Combining the above results together, Eq. \eqref{contractei} gives the following identity:
\ba\begin{aligned}\label{eq1}
H_{ab}\xi^a\gamma_c{}^b=-\left[1-\frac{\alpha}{4}(D\chi)^2\right]D_c\k-\frac{\a}{2}D^a\c D_c\c D_a\k\,.
\end{aligned}\ea
Considering {the condition \eq{domT} which is given by the dominant energy condition}, we have
\ba
\left[1-\frac{\alpha}{4}(D\chi)^2\right]D_c\k+\frac{\a}{2}D^a\c D_c\c D_a\k=0\,.
\ea

\subsection{Generalized zeroth law}
In this subsection, we would like to give a discussion to the above results. As suggested by Ghosh and Sarkar\cite{Ghosh:2020dkk}, we rewrite the identity \eqref{eq1} to a different form
\be\begin{aligned}\label{eq2}
X_c=\alpha M^a_c X_a\,,
\end{aligned}\ee
where
\be\begin{aligned}
X_c\equiv D_c \kappa\,,
\end{aligned}\ee
and
\be\begin{aligned}\label{Mac}
  M_c{}^a\equiv\frac{1}{4}(D\c)^2\delta_c{}^a-\frac{1}{2}D_c\c D^a\c\,.
\end{aligned}\ee
Since we assume that all quantities have a smooth limit to general relativity when $\alpha\rightarrow 0$ in Horndeski gravity. With a similar argument to Ref. \cite{Ghosh:2020dkk} will arise, $X_a$ can be expanded in terms of $\alpha$ and can be written as
\be\begin{aligned}
X_a=(X_0)_a+\alpha(X_1)_a+\alpha^2(X_2)_a+\dots\,,
\end{aligned}\ee
where $(X_0)_a$ is the corresponding $X_a$ in Einstein gravity and therefore it is zero. Similarly, $M_c^a$ can also be written as
\be\begin{aligned}\label{expandM}
M_c^a=(M_0)_c^a+\alpha(M_1)_c^a+\alpha^2(M_2)_c^a+\dots\,.
\end{aligned}\ee
{According to the assumption that all quantities have a smooth limit to general relativity, the scalar field should vanish when $\alpha\rightarrow 0$. Then, a closer look at \eqref{Mac} will reveal that $M_a^c$ is a quantity of $\alpha^2$ order, i.e. $(M_0)_c^a=(M_1)_c^a=0$. However, as we will discuss below, we can expand $M_c^a$ as \eqref{expandM} since the actual order of $M_c^a$ does not affect the final result. Eq. \eqref{eq2} can be expanded in terms of $\alpha$}
\be\begin{aligned}
  (X_0)_c+\alpha(X_1)_c+\dots=&\alpha(M_0)_c^a(X_0)_a+\alpha^2[(M_0)_c^a(X_1)_a\\
  &+(M_1)_c^a(X_0)_a]+\dots\,.
\end{aligned}\ee
This identity is established for any $\alpha$, and every order of $\alpha$ should be equal on both sides. {At first, we have $(X_0)_a=0$. Then, we equal the first order of $\alpha$. With the face that both $(M_0)_c^a$ and $(X_0)_a$ are zero, we have $(X_1)_c=(M_0)_c^a(X_0)_a=0$. For the second order, we have $(X_2)_c=(M_0)_c^a(X_1)_a+(M_1)_c^a(X_0)_a$, since $(X_0)_a$ and $(X_1)_a$ are zero, we can find $(X_2)_a=0$. For the $k$th order of $\alpha$, the identity gives
\be\begin{aligned}
(X_k)_c=\sum_{i=0}^{k-1} (M_i)_c^a(X_{k-i-1})_a\,.
\end{aligned}\ee
By a simple mathematical induction, we can prove that $(X_2)_c=(X_3)_c=\dots=0$ one by one.} And finally, we can get $X_c=0$. As the discussion above, the zeroth law always holds as long as $X_c$ satisfies the equation with the same form as \eqref{eq2} once we throw the smooth condition to the gravitational theory. And the zeroth law holds in Horndeski gravity.

\section{conclusion}\label{sec5}
In this paper, we studied the zeroth law in Horndeski gravity. We first reviewed the proof of the black hole zeroth law in Einstein gravity and gave some basic setups on the Killing horizon. Then, according to a similar calculation as Ref. \cite{Ghosh:2020dkk}, we calculated the derivative of the surface gravity $\kappa$ in each direction along the event horizon in the stationary black holes based on the equation of motion in Horndeski gravity {with the assumption that the scalar field $\chi$ shares the same symmetries with the spacetime}. We found that the derivative of $\kappa$ along the transverse directions satisfies such an identity
\be\begin{aligned}
D_c\kappa=\alpha M_c^a D_a\k\,
\end{aligned}\ee
once we assume the minimally coupled matter fields satisfy the dominant energy condition. With a similar discussion to Ref. \cite{Ghosh:2020dkk}, according to the above identity, we can get $D_a\kappa=0$ as long as all the quantities have a smooth limit to Einstein gravity. Then, combining with the universal relation $\xi^a\nabla_a\kappa=0$, this result means that the surface gravity $\kappa$ is a constant on the whole stationary horizon in Horndeski gravity, and therefore the zeroth law is satisfied.

{It is worth noting that we only studied a part of the Horndeski gravity theory. The zeroth law of the full Horndeski theory is also worth studying. The tedious equation of motion of the full Horndeski theory is calculated in Ref. \cite{Kobayashi:2011nu} and we can discover the terms appearing in the equation are quite similar with \eqref{horneom}. The further extension of the zeroth law proof to general Horndeski gravity theory will be complex but not hard.
Besides, in our paper, we assumed that the scalar field $\chi$ also shares the Killing symmetries. However, in Horndeski gravity, there exists the solution where $\math{L}_{\xi}\chi\neq 0$. Solutions like this is found in Refs. \cite{Babichev:2013cya,Babichev:2016rlq}. The spacetime in this solution is spherically symmetric while the scalar field is given by $\chi(t,r)=qt+\psi(r)$. It is obvious that for this solution, the scalar field $\chi$ does not satisfy $\math{L}_{\xi}\chi\neq 0$. It is also very interesting to further consider how to prove the zeroth law in this kind of solution.}

\section*{Acknowledgement}
This research was supported by National Natural Science Foundation of China (NSFC) with Grants No. 11775022 and 11873044.

\end{document}